# Randomized Ternary Search Tries

Nicolai Diethelm

Abstract

This paper presents a new kind of self-balancing ternary search trie that uses a randomized balancing strategy adapted from Aragon and Seidel's randomized binary search trees ("treaps"). After any sequence of insertions and deletions of strings, the tree looks like a ternary search trie built by inserting strings in random order. As a result, the time cost of searching, inserting, or deleting a string of length $k$ in a tree with $n$ strings is at most $O(k + \log n)$ with high probability.

## 1. Introduction

Tries are an important data structure for storing a set of character strings over a given alphabet. They can be used for a wide variety of search problems, from simple membership queries to more complex problems such as finding all strings that approximately match a given string.

The standard trie for a set $S$ of $n$ strings can be defined to be the smallest tree such that each node except the root is labeled with a character, the children of a node are alphabetically ordered, $n$ nodes are marked as a terminal node, and each string in $S$ is spelled out by a path from the root to a terminal node. Each node of the trie corresponds to a prefix of a string in $S$, namely the prefix spelled out by the path from the root to that node.

In a standard trie, a node can have as many children as there are characters in the alphabet. But if we organize the children of a standard trie node as a binary search tree, we get a trie structure with at most three children per node – a left child and a right child as in a binary search tree, and a middle child which is the root of the next binary search tree. The advantage of such a so-called "ternary search trie" is that it allows fast searches and updates without being wasteful of space or difficult to implement (see Bentley and Sedgewick [2]).

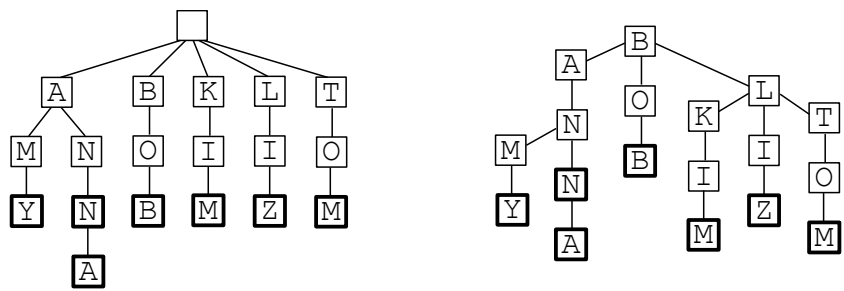

**Figure 1.** A standard trie (left) and a ternary search trie (right), each storing the strings AMY, ANN, ANNA, BOB, KIM, LIZ, and TOM. The bold squares denote terminal nodes.

A search in a ternary search trie compares the current character in the search string with the character at the current node. If the search character is less, the search goes to the left child; if the search character is greater, the search goes to the right child; and if the search character is equal, the search goes to the middle child and proceeds to the next character in the search string. So, if a ternary search trie for $n$ strings is balanced such that each step to a left or right child cuts the number of terminal nodes in the subtree at least in half, then searching for a given string of length $k$ takes at most $O(k + \log n)$ time.

The algorithm for inserting a new string into a ternary search trie works analogously to the algorithm for inserting a new key into a binary search tree: it searches for the string in the tree, adding missing nodes at the point where the search runs off the end of the tree. Like a binary search tree, a ternary search trie is thus sensitive to insertion order. If the strings are inserted in lexicographic order, each of the binary search trees within a ternary search trie degenerates into a linked list, significantly increasing the cost of searches. Fortunately, also like a binary search tree, a ternary search trie can be rebalanced using rotations. A rotation reverses the parent–child relationship between two nodes in a binary search tree without affecting the tree's order property.

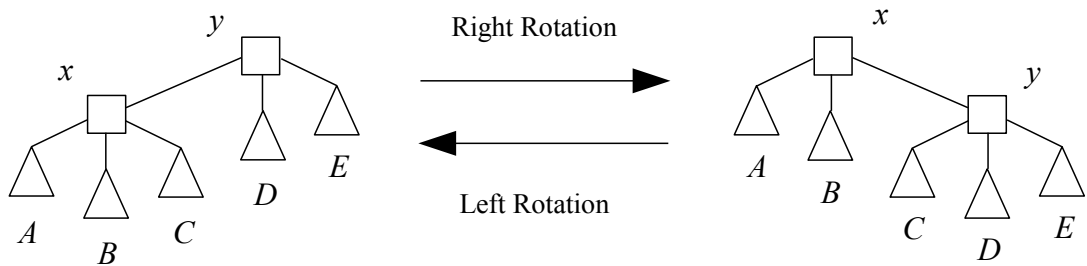

**Figure 2.** Rotations in a ternary search trie. The triangles represent subtrees.

Several known strategies for rebalancing a binary search tree via rotations have already been adapted for ternary search tries. Examples of such self-balancing ternary search tries are Mehlhorn's weight-balanced ternary search tries [3], Vaishnavi's multidimensional AVL trees [7], or Sleator and Tarjan's lexicographic splay trees [6]. This paper now presents a new kind of self-balancing ternary search trie that uses a randomized balancing strategy adapted from Aragon and Seidel's randomized binary search trees ("treaps") [5]. We begin with a definition of the new data structure.

## 2. Definition

Let $r$ be a positive integer. We define an "$r$-trie" for a set $S$ of strings to be a ternary search trie for $S$ with the following three properties:

1. Each string in $S$ has a random, uniformly distributed integer priority between 1 and $r$, with a higher number meaning a higher priority.
2. Each node has the priority of the highest-priority string that starts with the prefix corresponding to that node.
3. No node has a lower priority than its left or right child. In other words, each of the binary search trees within the ternary search trie is a heap with respect to the node priorities.

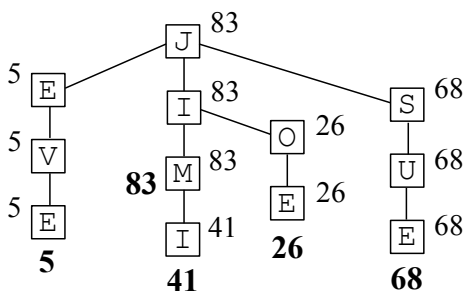

**Figure 3.** A 100-trie for the strings EVE, JIM, JIMI, JOE, and SUE. The bold numbers are the priorities of the strings. The other numbers are node priorities.

Properties 2 and 3 ensure that an *r*-trie is shaped like a ternary search trie built by inserting the strings in order of decreasing priority. For *r* → ∞, the shape of an *r*-trie becomes thus a random variable with the same probability distribution as the ternary search trie resulting from inserting the strings in random order.

An *r*-trie node has fields *char* (character), *prio* (priority of the node), *strPrio* (priority of the string corresponding to the node), *left*, *mid*, and *right* (pointers to the children). For a nonterminal node, *strPrio* is 0. Nonexistent child nodes (or an empty tree) are represented by the *nil* node, a sentinel node with priority 0.

## 3. Algorithms

The search algorithms for *r*-tries are the usual ones for ternary search tries. But how about updates? The insertion of a new string *s* into an *r*-trie can be achieved as follows:

Insert *s* with priority 0 using the usual insertion algorithm for ternary search tries.
Set the *strPrio* field of the terminal node of *s* to a random integer between 1 and *r*.
On the path from the terminal node to the root, for each node *x* that corresponds to a prefix of *s*:
    Set *x.prio* to max(*x.strPrio*, *x.mid.prio*).
        While *x* is the left or right child of its parent and has a higher priority than its parent:
            Rotate *x* with its parent.

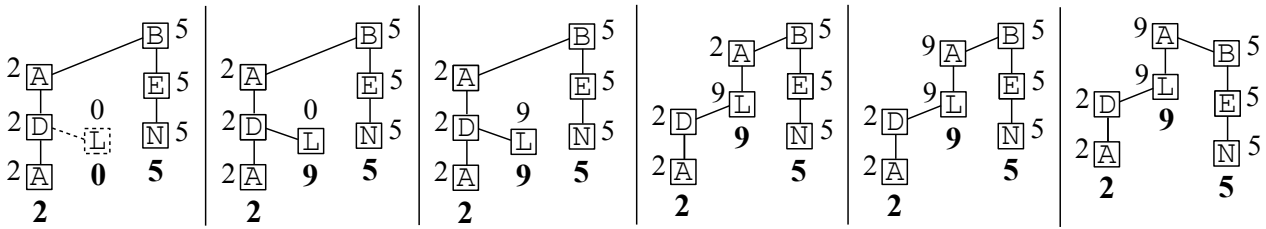

**Figure 4.** Insertion of string AL into a 10-trie.

The deletion of a string *s* from an *r*-trie can be achieved by "inverting" the insertion operation:

Set the *strPrio* field of the terminal node of *s* to 0.
On the path from the terminal node to the root, for each node *x* that corresponds to a prefix of *s*:
    Set *x.prio* to max(*x.strPrio*, *x.mid.prio*).
        While *x* has a lower priority than its left or right child:
             Rotate *x* with the higher-priority child.
        If *x* has priority 0, then unlink *x* from the tree (that is, replace *x* by the *nil* node).

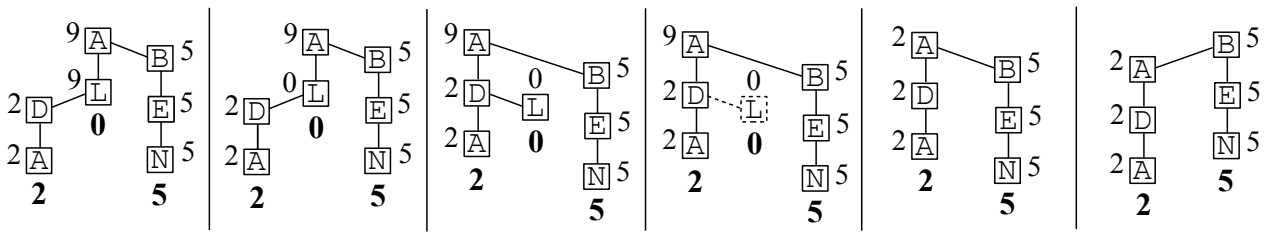

**Figure 5.** Deletion of string AL from a 10-trie.

## 4. Implementation

A recursive pseudocode implementation of the update algorithms for *r*-tries is shown below. The code inserts/deletes a given string *s* into/from an *r*-trie with the specified root. The expression *s*[*i*] denotes the *i*-th character of *s*.

```
root ← insert(s, 1, root)

insert(s, i, x):
    IF x = nil
        x ← newNode()
        x.char ← s[i]
        x.prio ← 0,  x.strPrio ← 0
        x.left ← nil,  x.mid ← nil,  x.right ← nil
    IF s[i] < x.char
        x.left ← insert(s, i, x.left)
        IF x.left.prio > x.prio
            x ← rotateWithLeft(x)
    ELSE IF s[i] > x.char
        x.right ← insert(s, i, x.right)
        IF x.right.prio > x.prio
            x ← rotateWithRight(x)
    ELSE
        IF i < s.length
            x.mid ← insert(s, i + 1, x.mid)
        ELSE IF x.strPrio = 0
            x.strPrio ← randomInteger(1, r)
        x.prio ← max(x.strPrio, x.mid.prio)
    RETURN x
```

```
root ← delete(s, 1, root)

delete(s, i, x):
    IF x ≠ nil
        IF s[i] < x.char
            x.left ← delete(s, i, x.left)
        ELSE IF s[i] > x.char
            x.right ← delete(s, i, x.right)
        ELSE
            IF i < s.length
                x.mid ← delete(s, i + 1, x.mid)
            ELSE
                x.strPrio ← 0
            x.prio ← max(x.strPrio, x.mid.prio)
            x ← heapifyOrDelete(x)
    RETURN x

heapifyOrDelete(x):
    IF x.prio < x.left.prio OR x.prio < x.right.prio
        IF x.left.prio > x.right.prio
            x ← rotateWithLeft(x)
            x.right ← heapifyOrDelete(x.right)
        ELSE
            x ← rotateWithRight(x)
            x.left ← heapifyOrDelete(x.left)
    ELSE IF x.prio = 0
        x ← nil
    RETURN x
```

```
rotateWithLeft(x):
    y ← x.left
    x.left ← y.right
    y.right ← x
    RETURN y
```

```
rotateWithRight(x):
    y ← x.right
    x.right ← y.left
    y.left ← x
    RETURN y
```

Function **insert**(*s*, *i*, *x*) inserts string *s* from its *i*-th character on into the subtree rooted at node *x*, returning the root of the updated subtree. The function compares the *i*-th character of *s* with the character at *x*, and then recursively calls itself on the left, right, or middle subtree of *x*. If the function runs off the end of the tree, it creates a new node, initializes the node, and falls through to the standard case. When the terminal node of *s* is reached, a random priority is generated and assigned to the string – but only if *s* is a new string in the tree. Finally, when the recursive calls return and the nodes are revisited in reverse order, the code for the node priority updates and the rotations is executed.

Function **delete**(*s*, *i*, *x*) works in a similar way. The subroutine **heapifyOrDelete**(*x*) rotates node *x* to its proper position in the heap, and then replaces *x* by the *nil* node if *x* has priority 0.

## 5. Analysis

A binary search tree built by inserting *n* keys in random order is known to have $O(\log n)$ height with high probability (see [4]). But how about the height of a randomized ternary search trie? To answer this question, let us first recall a simple fact about the ancestor relation in binary search trees:

**Lemma 1.** Let *x* and *y* be two keys in a binary search tree built by successively inserting keys. Then the node containing *x* is an ancestor of the node containing *y* if and only if *x* was inserted into the tree before *y* and before all keys that are greater than $\min(x, y)$ and less than $\max(x, y)$.

**Proof.** Clearly the lemma is true if *x* or *y* is at the root of the tree, or if *x* is in the left subtree of the root and *y* is in the right subtree of the root (or vice versa). And since the subtrees are binary search trees themselves, we get by induction that the lemma is also true if *x* and *y* are both in the left subtree or both in the right subtree. ∎

Now we can show an interesting relationship between ternary search tries and binary search trees:

**Lemma 2.** Let *S* be a set of strings, let *σ* be a permutation of *S*, and let $T_\sigma$ and $B_\sigma$ be the ternary search trie and binary search tree resulting from successively inserting the strings of *σ*. Then for each string *s* in *S*, searching for *s* in $T_\sigma$ requires at most as many "sidesteps" (steps to a left or right child) as searching for *s* in $B_\sigma$.

**Proof.** Let *P* be the path from the root of $T_\sigma$ to the terminal node of string *s*, and let $x_1, \ldots, x_m$ be the nodes on *P* whose left or right child belongs to *P* too (we assume there is at least one such node). Further, for $i = 1, \ldots, m$, let $t_i$ be the string in *S* whose insertion into the ternary search trie has created $x_i$. Then by applying Lemma 1 to the binary search tree containing $x_i$ we get that $t_i$ was inserted into the ternary search trie before *s* and before all strings that are lexicographically between $t_i$ and *s*. Since the same insertion order is true for $B_\sigma$, it follows from Lemma 1 again that in $B_\sigma$ the node containing $t_i$ is an ancestor of the node containing *s*. Consequently, because $t_1, \ldots, t_m$ are distinct, the node containing *s* has at least *m* ancestor nodes in $B_\sigma$. ∎

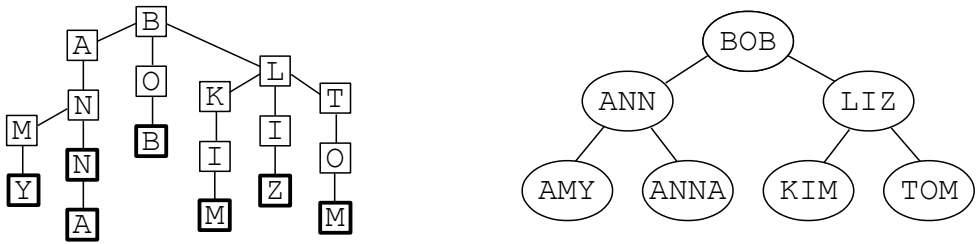

**Figure 6.** The ternary search trie and binary search tree resulting from successively inserting the strings BOB, ANN, AMY, ANNA, LIZ, KIM, and TOM.

Using Lemma 2, we get the following result about the height of an *r*-trie:

**Theorem 1.** For large enough *r*, the depth of the terminal node of a string of length *k* in an *r*-trie for *n* strings is at most $k + O(\log n)$ with high probability.

How about the cost of the basic operations on *r*-tries? A search for a string follows a path from the root to a terminal node until the search terminates. The insertion of a new string begins with a search for the string and requires at most as many rotations as there are edges to left and right child nodes along the search path. A deletion requires essentially the same number of rotations as an insertion since the deletion algorithm basically reverses an insertion operation. So we get:

**Theorem 2.** For large enough *r*, the time cost of searching, inserting, or deleting a string of length *k* in an *r*-trie for *n* strings is at most $O(k + \log n)$ with high probability.

## 6. Related Work

Badr and Oommen [1] described a ternary search trie whose binary search trees are randomized. In a randomized binary search tree, each node is equally likely to be the root. This is in contrast to a binary search tree within an *r*-trie, where nodes that correspond to a prefix of many strings tend to have a higher priority. A ternary search trie with randomized binary search trees is thus slower than an *r*-trie. It only provides a high-probability bound of $k \cdot O(\log d)$ on the depth of the terminal node of a string of length $k$, with $d$ being the size of the alphabet.